\documentclass{JHEP3}
\usepackage{amssymb,amsmath}
\usepackage{epsfig}
\usepackage{graphicx}
\usepackage{citesort}

\title{On the role of shear in cosmological averaging}

\author{Maria Mattsson$^{1,2,}$\footnote{E-mail: maria.ronkainen@helsinki.fi}~, Teppo Mattsson$^{2,}$\footnote{E-mail: teppo.mattsson@canterbury.ac.nz}\\
${}^1$ Physics Department and Helsinki Institute of Physics, P.O.Box 64, FIN-00014 University of Helsinki, Finland\\
${}^2$ Department of Physics and Astronomy, University of
Canterbury, Private Bag 4800, Christchurch 8140, New Zealand\\}

\abstract{Using the spherically symmetric inhomogeneous
Lema\^itre-Tolman-Bondi dust solution, we study how the shear and
the backreaction depend on the sharpness of the spatial transition
between voids and walls and on the size of the voids. The voids
considered here are regions with matter density $\Omega_0 \simeq
0$ and expansion rate $H_0 t_0 \simeq 1$, while the walls are
regions with matter density $\Omega_0 \simeq 1$ and expansion rate
$H_0 t_0 \simeq 2/3$. The results indicate that both the
volume-average shear and the variance of the expansion rate grow
proportional to the sharpness of the transition and diverge in the
limit of a step function, but, for realistic-sized voids, are
virtually independent of the size of the void. However, the
backreaction, given by the difference of the variance and the
shear, has a finite value in the step-function limit. By comparing
the exact result for the backreaction to the case where the shear
is neglected by treating the voids and walls as separate
Friedmann-Robertson-Walker models, we find that the shear
suppresses the backreaction by a factor of $(r_0/t_0)^2$, the
squared ratio of the void size to the horizon size. This
exemplifies the importance of using the exact solution for the
interface between the regions of different expansion rates and
densities. The suppression is justified to hold also for a network
of compensated voids, but may not hold if the universe is
dominated by uncompensated voids.}

\preprint{HIP-2010-19/TH}

\keywords{Inhomogeneous Cosmological Models, Averaging in General
Relativity, Cosmology, Gravitation}

\begin{document}

\section{Introduction}\label{intro}

An unresolved issue in cosmology is the effect of the structure
formation on the cosmological observations beyond perturbative
analysis of the Friedmann-Robertson-Walker or FRW models
\cite{Ell84,EllisStoeger,Ellis:1999sx,Rasanen:2010wz}. On account
of the increased precision of these observations and the cosmic
acceleration that they seem to indicate
\cite{Riess:2006fw,Percival:2009xn,Komatsu:2010fb}, the evaluation
of this effect has become important
\cite{Sarkar:2007cx,Buchert:2007ik,Mattsson:2007tj,Wiltshire:2007zj}.
A way to estimate the effect of the cosmic structures is via a
backreaction term that arises by averaging inhomogeneous scalar
quantities on spatial hypersurfaces \cite{Buchert:1999er}. The
backreaction thus obtained is given by the variance of the
expansion rate minus the non-negative average shear.

A simplification in some estimates of the cosmological
backreaction is based on partially or fully neglecting the shear
on the interface between regions of different expansion rates
\cite{Rasanen:2006kp,Rasanen:2008it,Wiltshire:2007jk}: these
studies have found a significant amount of backreaction. On the
other hand, perturbative studies that do not neglect the shear
have suggested the backreaction to be insignificant
\cite{Gruzinov:2006nk,Paranjape:2008jc,Brown:2008ra,Baumann:2010tm};
however, see Ref.\ \cite{Rasanen:2010wz} for a recent discussion
on the possible shortcomings of the perturbative approach in the
backreaction problem. In this paper, we consider the issue within
exact general relativity by studying:
\begin{enumerate}

\item How the sharpness of the spatial transition between the
voids (where $\Omega_0 \simeq 0$ and $H_0 t_0 \simeq 1$) and the
walls (where $\Omega_0 \simeq 1$ and $H_0 t_0 \simeq 2/3$), as
well as the size of the voids, affect the cosmic shear and thus
the backreaction in the spherically symmetric inhomogeneous
Lema\^itre-Tolman-Bondi or LTB dust models.

\item How the exact backreaction of the void-wall configuration
relates to the case where the shear is neglected by treating the
voids and the walls as separate FRW solutions.

\end{enumerate}
We perform the calculations on a spatial hypersurface at $t=t_0$
and do not explicitly consider time-dependence (though $t_0$ can
be considered arbitrary).

The systematic study on the role of shear makes our approach
different from the previous studies on the dynamical backreaction
in the LTB model which have focused on finding profiles that
exhibit acceleration of the average expansion
\cite{Chuang:2005yi,Paranjape:2006cd} or on general properties of
the backreaction \cite{Sussman:2008xp,Sussman:2008vs}.

The paper is organized as follows. The necessary background of the
LTB solution and the Buchert averaging method are introduced in
Sects.\ \ref{ELTEEBEE} and \ref{averaging}, respectively. In
Sect.\ \ref{sier}, we consider the shear and, in Sect.\
\ref{Sbackreaction}, the backreaction for almost compensated LTB
voids residing within almost FRW walls. To show that the results
do not depend on the compensating overdensity, we discuss
uncompensated voids with a monotonically increasing physical
matter density profile in Sect.\ \ref{uncompensated}. Finally, the
conclusions are given in Sect.\ \ref{konkluusiot}.

\section{LTB solution}\label{ELTEEBEE}

The exact spherically symmetric dust solution of general
relativity was discovered by Lema\^itre in 1933
\cite{Lemaitre:1933qe} and is now commonly known as the LTB
metric:
\begin{equation}\label{LTBmetric}
{\rm d}s^2 = - {\rm d}t^2 + \frac{[A'(r,t)]^2}{1-k(r)}{\rm d}r^2 +
A^2(r,t) ({\rm d} \theta^2 + \sin^2 \theta {\hspace{1pt}} {\rm d}
\varphi^2)~,
\end{equation}
where $A'(r,t) \equiv \partial_r A(r,t)$, $k(r)$ is related to the
curvature of the spatial sections and $A(r,t)$ is determined by
the Friedmann-like evolution equation \cite{Enqvist:2006cg}
\begin{equation}\label{LTBfriedman}
H(r,t) = H_0(r) \left[ \Omega_0(r) \left(\frac{A_0(r)}{A(r,t)}
\right)^3 + (1- \Omega_0(r)) \left(\frac{A_0(r)}{A(r,t)} \right)^2
\right]^{1/2}~,
\end{equation}
where $H(r,t) \equiv \partial_t A(r,t)/A(r,t) \equiv
\dot{A}(r,t)/A(r,t)$, $H_0(r) \equiv H(r,t_0)$ and $\Omega_0(r)$
are boundary condition functions specified on a spatial
hypersurface $t=t_0$ that determine the radial inhomogeneity
profile, while the freedom to choose the function $A_0(r) \equiv
A(r,t_0)$ corresponds to the scaling of the $r$-coordinate, used
in this work to set
\begin{equation}\label{Coordinates}
A_0(r) = r~.
\end{equation}
The curvature function $k(r)$ in the metric (\ref{LTBmetric}) is
related to these by
\begin{equation}\label{kcurvature}
k(r) \equiv H_0^2(r) (\Omega_0(r) - 1) A_0^2(r)~,
\end{equation}
and the boundary condition function $\Omega_0(r)$ is related to
the physical matter density $\rho(r,t)$ on the $t=t_0$
hypersurface as
\begin{equation}\label{LTBomega}
\Omega_0(r) \equiv \frac{8 \pi G}{3 H_0^2 (r)}
\frac{\int_{\mathbb{B}(r)} \rho_0(r) {\rm d}^3
x}{\int_{\mathbb{B}(r)} {\rm d}^3 x}~,
\end{equation}
where $\rho_0(r) \equiv \rho(r,t_0)$. Inversely, $\rho_0(r)$ can
be written in terms of $\Omega_0(r)$ and $H_0(r)$ as
\begin{equation}\label{rooOomega}
\rho_0(r) =  \frac{3 H_0^2(r)}{8 \pi G} \Omega_0(r) \left[1 +
\frac{A_0(r)}{3 A_0'(r)} \left( \frac{\Omega_0'(r)}{\Omega_0(r)}
+ 2 \frac{H_0'(r)}{H_0(r)} \right) \right]~.
\end{equation}

We only consider LTB models where the boundary condition functions
obey the constraint
\begin{equation}\label{H0-OmegaRelationAppr}
H_0(r)=\frac{1}{t_0}\left(1-\frac{\sqrt{\Omega_0(r)}}{3}\right)~,
\end{equation}
which approximates the simultaneous Big Bang condition
\begin{equation}\label{H0-OmegaRelationExact}
H_0(r)=\frac{1}{t_0}\left[\frac{\sqrt{1-\Omega_0(r)}-\Omega_0(r)\rm{arsinh}\sqrt{\frac{1-\Omega_0(r)}{\Omega_0(r)}}}{(1-\Omega_0(r))^{3/2}}\right]~¨,
\end{equation}
such that the $|\rm{error}|<1.5\%$ in the considered interval $0
\le \Omega_0 \le 1$ and no error at the extremes $\Omega_0=0$ and
$\Omega_0=1$. We use Eq.\ (\ref{H0-OmegaRelationAppr}) instead of
Eq.\ (\ref{H0-OmegaRelationExact}) to make analytic calculations
possible. The models with (approximately) simultaneous Big Bang
form perhaps the most relevant subcase of LTB solutions, because,
in these models, the inhomogeneities are growing modes (see
\cite{Silk}) as e.g.\ the near isotropy of the CMB suggests is
also the case in the real universe.

To study the shear in Sect.\ \ref{sier} and the backreaction in
Sect.\ \ref{Sbackreaction}, we need the following quantities of
the LTB model: the shear scalar
\begin{equation}\label{LTBshear1}
\sigma^2(r,t) \equiv\sigma^{\mu\nu}\sigma_{\mu\nu} = \frac{2}{3}
\left( \frac{\dot{A}(r,t)}{A(r,t)} - \frac{\dot{A}'(r,t)}{A'(r,t)}
\right)^2~,
\end{equation}
the volume expansion scalar
\begin{equation}\label{LTBexpansion}
\theta(r,t) = 2 \frac{\dot{A}(r,t)}{A(r,t)} +
\frac{\dot{A}'(r,t)}{A'(r,t)} = \frac{1}{A'(r,t) A^2(r,t)}
\frac{\partial}{\partial r} \left( A^2(r,t) \dot{A}(r,t) \right)
~,
\end{equation}
and their expressions on the $t=t_0$ hypersurface:
\begin{equation}\label{LTBshear2}
\sigma^2 (r,t_0) = \frac{2}{3}\left(rH'_0(r)\right)^2~,
\end{equation}
\begin{equation}\label{LTBtheta}
\theta(r,t_0) =
3H_0(r)+rH'_0(r)=\frac{1}{r^2}\frac{\partial}{\partial
r}\left(r^3H_0(r)\right)~.
\end{equation}
Using the relation (\ref{H0-OmegaRelationAppr}), the shear
(\ref{LTBshear2}) can also be expressed in terms of $\Omega_0(r)$
as
\begin{equation}\label{LTBshear3}
\sigma^2 (r,t_0) =
\frac{1}{54}\left(\frac{r}{t_0}\right)^2\frac{\left(\Omega_0'(r)\right)^2}{\Omega_0(r)}~.
\end{equation}
We also have use for the LTB volume element
\begin{equation}\label{LTBvolume}
\sqrt{g} {\hspace{1pt}} {\rm d}r {\hspace{1pt}} {\rm d}\theta
{\hspace{1pt}} {\rm d}\varphi = \frac{A'(r,t)A^2(r,t) \sin
\theta}{\sqrt{1-k(r)}} {\hspace{1pt}} {\rm d}r {\hspace{1pt}} {\rm
d}\theta {\hspace{1pt}} {\rm d}\varphi~,
\end{equation}
where $g$ denotes the determinant of the spatial part of the
metric.

\subsection{The void-wall model}\label{model}

We use the LTB solution to model a configuration that consists of
two different regions: a void where $\Omega_0 \simeq 0$ and a wall
where $\Omega_0 \simeq 1$, with a smooth transition in between.
For this we choose the $\Omega_0(r)$ profile as follows:
\begin{equation}\label{OmegaLTB}
\Omega_0(r)=\left(1-e^{-({r}/{r_0})^n}\right)^2~,
\end{equation}
where $r_0$ determines the size of the void and $n$ the sharpness
of the spatial transition between the void and the wall. By virtue
of the simultaneous Big Bang condition
(\ref{H0-OmegaRelationAppr}), Eq.\ (\ref{OmegaLTB}) implies
\begin{equation}\label{H0LTB}
H_0(r) = \frac{t_0^{-1}}{3} \left( 2  +  e^{-({r}/{r_0})^n}
\right)~,
\end{equation}
which tells us that the void expands faster than the wall by a
factor of $3/2$.

Using LTB models with the void-wall profile (\ref{OmegaLTB}), we
can easily join together many voids to construct a model for a
\emph{network} of voids. This is because LTB solutions with the
profile (\ref{OmegaLTB}) are, up to negligible terms of order
$e^{-(R/r_0)^n}$, $\Omega=1$ FRW dust solutions outside the void
at $R>r_0$, all with $t_0$ as the age of the universe, implying
that the different LTB solutions naturally join together in the
wall region.

\subsection{Small $k(r)$ expansion}

For a realistic or sub-horizon size void $r_0 \ll t_0$, which we
next show implies $| k(r) | \ll 1 ~ \forall ~ r$. Note, however,
that since the Ricci scalar of the spatial sections is given by
\begin{equation}\label{3RicciScalar}
{\mathcal{R}} = \frac{2}{A^2(r,t) A'(r,t)}
\frac{\partial}{\partial r} \left( A(r,t) k(r) \right)~,
\end{equation}
or at $t=t_0$ by
\begin{equation}\label{3RicciScalart0}
{\mathcal{R}}_0 = \frac{2}{r^2} \frac{\partial}{\partial r}
\left(r k(r) \right),
\end{equation}
the condition $| k(r) | \ll 1$ does not imply that the spatial
curvature is small in the void.

With the conditions (\ref{Coordinates}) and
(\ref{H0-OmegaRelationAppr}), the curvature function
(\ref{kcurvature}) becomes
\begin{equation}\label{kcurvature2}
k(r) = - \left( \frac{r}{t_0} \right)^2 \underbrace{\left(1 -
\frac{1}{3} \sqrt{\Omega_0(r)} \right)^2 \left(1- \Omega_0(r)
\right)}_{ \leq 1}~.
\end{equation}
For the considered profiles, $\Omega_0(r)$ rapidly approaches the
value $\Omega_0=1$ outside the void or when $r>r_0$, thus implying
\begin{equation}\label{kcurvature3}
{\rm{max}}\left(|k(r)|\right) \simeq \left( \frac{r_0}{t_0}
\right)^2 \ll 1 ~.
\end{equation}

As is evident in Eqs.\ (\ref{kcurvature}) and (\ref{kcurvature2}),
$k(r)$ depends only on the inhomogeneity profile functions
$\Omega_0(r)$ and/or $H_0(r)$, but not on their derivatives. Hence
the small $k(r)$ approximation is equally valid for all the
profiles considered here regardless of the sharpness of the
transition between the void and the wall.

Since $|k(r)| \ll 1$, the part of the LTB volume element
(\ref{LTBvolume}) that contains the curvature function $k(r)$ can
be expanded as follows
\begin{equation}\label{curvatureExpanded}
\frac{1}{\sqrt{1-k(r)}} = 1 + \frac{1}{2}k(r) + \mathcal{O}
(k^2(r))~.
\end{equation}
Furthermore, Eqs.\ (\ref{LTBmetric}), (\ref{Coordinates}) and
(\ref{kcurvature3}) imply that the coordinate $r$ closely measures
the proper distance on the spatial hypersurface defined by
$t=t_0$.

\section{Scalar averaging}\label{averaging}

The spatial volume-average of a scalar $S(x,t)$ is defined as
\begin{equation}\label{defAverage}
\langle S(x,t) \rangle_{\mathcal{D}} \equiv
\frac{\int_{\mathcal{D}} S(x,t) \sqrt{g} {\hspace{1pt}} {\rm d}^3
x}{\int_{\mathcal{D}} \sqrt{g} {\hspace{1pt}} {\rm d}^3 x}~,
\end{equation}
where $g$ is the determinant of the spatial metric and
$\mathcal{D}$ is the averaging domain or a region of the spatial
sections.

By applying the averaging (\ref{defAverage}) to the scalar parts
of the Einstein equation in an irrotational dust universe, we
obtain the Buchert equations \cite{Buchert:1999er}:
\begin{eqnarray}
3\frac{\ddot{a}_\mathcal{D}(t)}{a_\mathcal{D}(t)} &=& -4 \pi G \langle\rho\rangle_\mathcal{D}(t)+\mathcal{Q}_\mathcal{D}(t)~ \label{BuchertFirst2} \\
3\left(\frac{\dot{a}_\mathcal{D}(t)}{a_\mathcal{D}(t)}\right)^2
&=& 8 \pi G\langle\rho\rangle_\mathcal{D}(t)-\frac{1}{2}\langle
{\mathcal{R}}
\rangle_\mathcal{D}(t)-\frac{1}{2}\mathcal{Q}_\mathcal{D}(t)~
\label{BuchertSecond2} \\
\partial_t\langle\rho\rangle_\mathcal{D}(t)&=&-3\frac{\dot{a}_\mathcal{D}(t)}{a_\mathcal{D}(t)}\langle\rho\rangle_\mathcal{D}(t)~,
\label{BuchertThird2}
\end{eqnarray}
where the regional scale factor is defined as
\begin{equation}\label{ExtendedScaleFactor}
a_\mathcal{D}(t)\equiv\left(
\frac{\int_{\mathcal{D}}\sqrt{g(x^i,t)}{\hspace{1pt}}{\rm
d}^3x}{\int_{\mathcal{D}}\sqrt{g(x^i,t_0)}{\hspace{1pt}}{\rm
d}^3x}\right)^{1/3}
\end{equation}
and the backreaction term
\begin{equation}\label{defBackreaction}
\mathcal{Q}_\mathcal{D}(t) \equiv \frac{2}{3}\left( \langle
\theta^2 \rangle_\mathcal{D} - \langle \theta
\rangle^2_\mathcal{D} \right) - \langle \sigma^2
\rangle_\mathcal{D}
\end{equation}
quantifies the difference of the time evolution of the averages
relative to the homogeneous quantities in an FRW dust universe. In
what follows, we take the averaging domain to be an
origin-centered ball of coordinate radius $R$, i.e.\
$\mathcal{D}=\mathbb{B}(R)$, and do not write it explicitly
anymore, so from here on for all scalars $S$:
\begin{equation}\label{dropoutdomain}
\langle S \rangle_{\mathcal{D}} = \langle S
\rangle_{\mathbb{B}(R)} = \langle S \rangle ~.
\end{equation}

\subsection{Averaging in the LTB model}

Given the LTB volume element (\ref{LTBvolume}), we can expand Eq.\
(\ref{defAverage}) for small $k(r)$ using the result
(\ref{curvatureExpanded}):
\begin{equation}\label{expandedLTBaverage}
\langle S \rangle = \langle S \rangle_0 + \frac{1}{2} \left(
\langle S k \rangle_0 - \langle S \rangle_0 \langle k \rangle_0
\right) + \mathcal{O}(k^2)~,
\end{equation}
where the subscript $0$ now refers to averages where $k(r)=0$ in
the integration measure, that is $\sqrt{g_0} \equiv
A'(r,t)A^2(r,t) \sin \theta$. Using this notation, the expression
of $\Omega_0(r)$ in Eq.\ (\ref{LTBomega}) can be recognized simply
as
\begin{equation}\label{LTBomega3}
\Omega_0(r) = \frac{\langle \rho_0(r) \rangle_0}{\rho_{{\rm
crit}}(r)} ~,
\end{equation}
where $\rho_{{\rm crit}}(r) \equiv 3 H_0^2(r)/8 \pi G$.

\section{Shear}\label{sier}

\begin{figure}[tbh]
\begin{center}
\includegraphics[width=10.0cm]{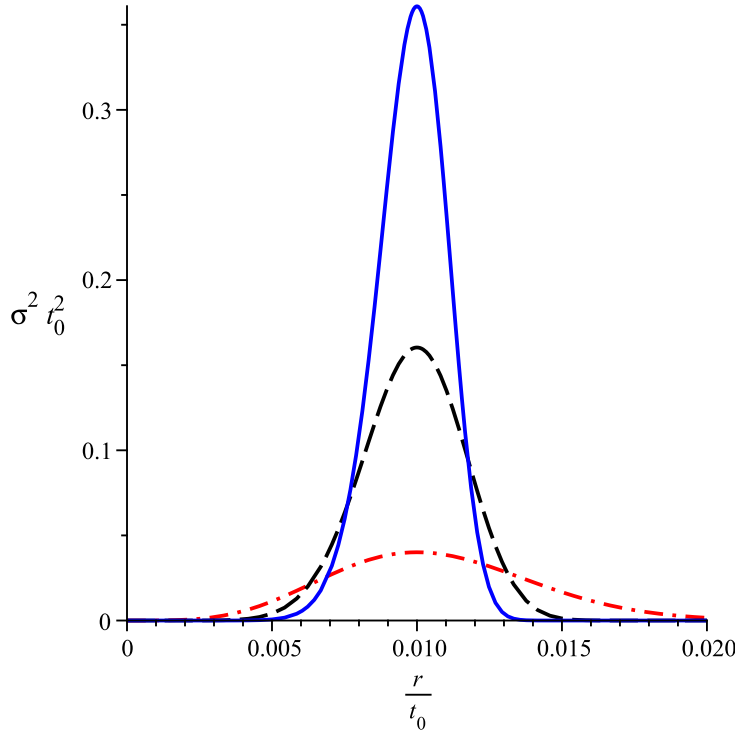}
\caption{Shear distribution as a function of $r$ for three
different values of $n$: $n=2$ (red dash dot curve), $n=4$ (black
dashed curve) and $n=6$ (blue solid curve). All profiles in this
figure have $r_0 = 0.01 t_0$.}\label{shear1}
\end{center}
\end{figure}

When the transition between the void and the wall is made sharper
by increasing $n$, the shear density on the interface grows but it
also becomes more localized. This can be seen in Eq.\
(\ref{LTBshear3}) and is illustrated in Fig.\ \ref{shear1}, where
we plot $\sigma^2(r,t_0)$ for a few profiles (\ref{OmegaLTB}) with
different values of $n$. It is hence a priori unclear how the
integrated or volume-averaged shear behaves when $n$ is varied.

\begin{figure}[tbh]
\begin{center}
\includegraphics[width=10.0cm]{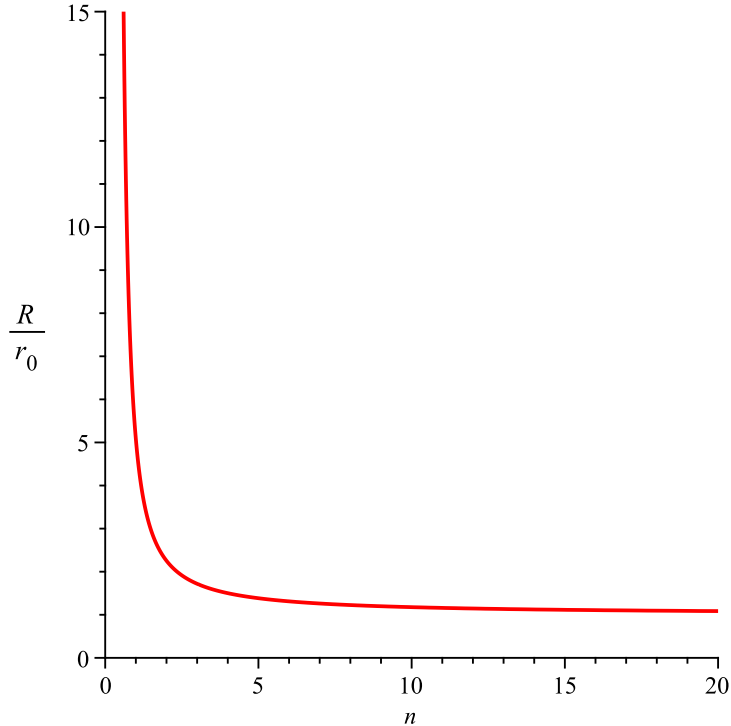}
\caption{The area above the curve (conservatively) represents the
region in the parameter space where the approximation ({\protect
\ref{shearapproximation}}) is valid.} \label{approxi1}
\end{center}
\end{figure}

We apply the expansion (\ref{expandedLTBaverage}) for the shear:
\begin{equation}\label{expandedLTBaverageshear}
\langle \sigma^2 \rangle = \langle \sigma^2 \rangle_0 +
\underbrace{\frac{1}{2} \left( \langle \sigma^2 k \rangle_0 -
\langle \sigma^2 \rangle_0 \langle k \rangle_0 \right)}_{\equiv
\langle \sigma^2 \rangle_1}~.
\end{equation}
Since the shear falls off rapidly in the wall outside the void, we
can also make the following approximation
\begin{equation}\label{shearapproximation}
\int_{0}^{R} \sigma^2(r,t_0) r^2 {\hspace{1pt}} {\rm d} r \simeq
\int_{0}^{\infty} \sigma^2(r,t_0) r^2 {\hspace{1pt}} {\rm d}r ~,
\end{equation}
where $R>r_0$ is the (coordinate) radius of the spherical
averaging region. The validity of the approximation
(\ref{shearapproximation}) depends on $n$: $n=1$ requires $R/r_0
\gtrsim 6$, $n=2$ requires $R/r_0 \gtrsim 2$, $n=6$ requires
$R/r_0 \gtrsim 1.2$, while for very large $n$ it is enough just to
have $R/r_0 > 1$; see Fig.\ \ref{approxi1}.

Using Eqs.\ (\ref{LTBshear3}), (\ref{LTBvolume}), (\ref{OmegaLTB})
and (\ref{defAverage}), along with the approximation
(\ref{shearapproximation}), we obtain the following analytic
expression for the average shear in the zeroth order of $k(r)$:
\begin{equation}\label{0thOrderLTBshear}
\langle\sigma^2\rangle_0 =
\frac{t_0^{-2}}{6}\left(\frac{r_0}{R}\right)^3\left(1+\frac{3}{n}\right)\Gamma
\left(\frac{3}{n}\right)2^{-3/n}~,
\end{equation}
where $\Gamma$ stands for Euler's gamma function. Similarly, the
first order term for shear is
\begin{eqnarray} \label{1stOrderLTBshear}
\langle\sigma^2\rangle_1 &=& t_0^{-2} \left(\frac{r_0}{R}\right)^3
\left( \frac{r_0}{t_0} \right)^2
\left\{\frac{5}{81}\left(1+\frac{5}{n}\right)
\Gamma\left(\frac{5}{n}\right)\left[- 8 \cdot
3^{-5/n-2}-4^{-5/n-1}+ 2 \cdot 5^{-5/n-2} +6^{-5/n - 2}\right]\right. \nonumber \\
&-&\left. \left(\frac{r_0}{R}\right)^3 \frac{1}{36n}
\left(1+\frac{3}{n}\right) \Gamma \left(\frac{3}{n}\right) \Gamma
\left(\frac{5}{n}\right)2^{-3/n}\left[-8-4\cdot 2^{-5/n}+2\cdot
3^{-5/n}+4^{-5/n}\right]\right\}~,
\end{eqnarray}
which is suppressed by the overall factor $(r_0/t_0)^2$ relative
to the leading order term (\ref{0thOrderLTBshear}). Therefore, as
can be easily verified by numerical methods, for the values of
$R/r_0$ and $n$ above the curve in Fig.\ \ref{approxi1}, Eq.\
(\ref{0thOrderLTBshear}) gives the correct result for the average
shear.

An immediate conclusion from the expression
(\ref{0thOrderLTBshear}) is that, for realistic or sub-horizon
size voids, the average shear is independent of the size of the
void (as long as the volume-ratio $(r_0/R)^3$ is kept fixed for
each void-wall pair). Since, as explained in Sect.\ \ref{model},
we can join together many LTB void-wall profiles (\ref{OmegaLTB})
to form a network of voids, Eq.\ (\ref{0thOrderLTBshear}) thus
also gives the average shear for the network of sub-horizon size
voids where each void can have a different value of $r_0$.

For a universe with half of the volume in voids\footnote{Order of
magnitude consistent with observed values quoted e.g. in Ref.\
\cite{Hoyle:2003hc}.}, $(r_0/R)^3=1/2$. Using this value and
setting $n=5$ in Eq.\ (\ref{0thOrderLTBshear}), we obtain
\begin{equation}\label{shearnumerical}
\langle \sigma^2(r,t_0) \rangle = 0.1 {\hspace{1pt}} t_0^{-2} ~.
\end{equation}

In the step function limit $n \rightarrow \infty$, Eq.\
(\ref{0thOrderLTBshear}) diverges with the following asymptotics:
\begin{equation}\label{0thOrderLTBshearLargen}
\langle \sigma^2(r,t_0) \rangle \sim  \frac{1}{18} t_0^{-2}
\left(\frac{r_0}{R}\right)^3  n~,
\end{equation}
telling us that the sharper the transition between the void and
the wall, the higher the value of the average shear.

\section{Backreaction}\label{Sbackreaction}

We consider here the backreaction (\ref{defBackreaction}) for LTB
models with a void-wall profile given by Eqs.\ (\ref{OmegaLTB})
and (\ref{H0LTB}).

\subsection{Backreaction in 0th order of $k(r)$}\label{backreaction0}

Let us first show that the backreaction (\ref{defBackreaction})
vanishes in the 0th order of $k(r)$. For this it is helpful to
notice that the following terms can be written as total
derivatives:
\begin{equation}\label{LTBQ1}
\frac{2}{3} \theta^2 - \sigma^2 = 2 \frac{\dot{A}^2}{A^2} + 4
\frac{\dot{A}\dot{A}'}{A A'} = \frac{2}{A^2 A'}
\frac{\partial}{\partial r} \left( \dot{A}^2 A \right)~,
\end{equation}
\begin{equation}\label{LTBQ2}
\theta = 2 \frac{\dot{A}}{A} + \frac{\dot{A}'}{A'} = \frac{1}{A^2
A'} \frac{\partial}{\partial r} \left( \dot{A} A^2 \right)~.
\end{equation}
Since the volume element (\ref{LTBvolume}) in the 0th order of
$k(r)$ is just
\begin{equation}\label{LTBQ3}
\sqrt{g_0} {\hspace{1pt}} {\rm d} ^3 x = A^2 A' \sin \theta
{\hspace{1pt}} {\rm d} r {\hspace{1pt}} {\rm d} \theta
{\hspace{1pt}} {\rm d} \varphi~,
\end{equation}
we have
\begin{equation}\label{LTBQ4}
\Big\langle \frac{2}{3} \theta^2 - \sigma^2 \Big\rangle_0 =
\frac{3}{A^3} 2 \dot{A}^2 A = 6 \frac{\dot{A}^2(R,t)}{A^2(R,t)}
\end{equation}
and
\begin{equation}\label{LTBQ5}
\frac{2}{3} \langle \theta \rangle_0^2 = \frac{2}{3} \left(
\frac{3}{A^3} A^2 \dot{A} \right)^2 = 6
\frac{\dot{A}^2(R,t)}{A^2(R,t)}~,
\end{equation}
so
\begin{equation}\label{LTBQ6}
\mathcal{Q}_0 = \frac{2}{3} \left( \langle \theta^2 \rangle_0 -
\langle \theta \rangle_0^2 \right) - \langle \sigma^2 \rangle_0 =
\Big\langle \frac{2}{3} \theta^2 - \sigma^2 \Big\rangle_0 -
\frac{2}{3} \langle \theta \rangle_0^2 = 0~.
\end{equation}

Given the zeroth order term of the backreaction (\ref{LTBQ6})
vanishes, the variance of the expansion rate must be equal to the
average shear (\ref{0thOrderLTBshear}) to the leading order, so we
have
\begin{equation}\label{variance}
\frac{2}{3} \left( \langle \theta^2 \rangle_0 - \langle \theta
\rangle_0^2 \right) =
\frac{t_0^{-2}}{6}\left(\frac{r_0}{R}\right)^3\left(1+\frac{3}{n}\right)\Gamma
\left(\frac{3}{n}\right)2^{-3/n}~.
\end{equation}

\subsection{Backreaction in 1st order of $k(r)$}\label{backreaction1}

As the average shear and the variance of the expansion rate cancel
exactly in the leading order of $k(r)$, we must go beyond the
zeroth order to obtain the leading order term of the backreaction.

Using the result (\ref{expandedLTBaverage}), the backreaction
(\ref{defBackreaction}) becomes
\begin{eqnarray}\label{1stOrderLTBbackreaction}\nonumber
\mathcal{Q} = \frac{2}{3} \left( \langle \theta^2 \rangle_0
-\langle \theta \rangle_0^2 \right) - \langle \sigma^2 \rangle_0 +
\langle k \rangle_0 \left( \frac{1}{3} \langle \theta \rangle_0^2
- \frac{1}{2} \left[ \frac{2}{3} (\langle \theta^2 \rangle_0 -
\langle \theta \rangle_0^2) - \langle \sigma^2 \rangle_0 \right]
\right) + \\
+ \Big\langle \left(\frac{\theta^2}{3} - \frac{\sigma^2}{2}
\right) k \Big\rangle_0 - \frac{2}{3} \langle \theta \rangle_0
\langle \theta k \rangle_0~,
\end{eqnarray}
where, according to the result (\ref{LTBQ6}), $\mathcal{Q}_0
\equiv \frac{2}{3} \left( \langle \theta^2 \rangle_0 - \langle
\theta \rangle_0^2 \right) - \langle \sigma^2 \rangle_0$ vanishes,
so Eq.\ (\ref{1stOrderLTBbackreaction}) reduces to
\begin{equation}\label{1stOrderLTBbackreactionb}
\mathcal{Q} = 3 H_0^2(R) \langle k \rangle_0 + \Big\langle
\left(\frac{\theta^2}{3} - \frac{\sigma^2}{2} \right) k
\Big\rangle_0 - 2 H_0(R) \langle \theta k \rangle_0~,
\end{equation}
where we have written $\langle \theta \rangle_0$ in terms of
$H_0(R)$ using the result (\ref{LTBQ5}).

Using the approximation (\ref{shearapproximation}) for the
integrals in Eq.\ (\ref{1stOrderLTBbackreactionb}) that contain
$k(r)$ as a common factor, we obtain
\begin{eqnarray} \label{1stOrderLTBbackreactionResult}
\mathcal{Q} &=& \frac{1}{9} t_0^{-2} \left( \frac{r_0}{R}
\right)^3 \left( \frac{r_0}{t_0} \right)^2  \frac{1}{n} \Gamma
\left( \frac{5}{n} \right) \left\{ \frac{8}{9} \cdot 3^{-5/n} -
\frac{2}{3} \cdot 4^{-5/n} +
\frac{2}{3} \cdot 5^{-5/n} + \frac{4}{9} \cdot 6^{-5/n}+  \right. \nonumber \\
&+& e^{- 2 (R/r_0)^n} \left( -8 - 4 \cdot 2^{-5/n} + 2 \cdot
3^{-5/n} + 4^{-5/n} \right) + \nonumber \\
&+& \left. e^{-(R/r_0)^n} \left( \frac{8}{3} \cdot 2^{-5/n} +
\frac{32}{9} \cdot 3^{-5/n} - \frac{7}{3} \cdot 4^{-5/n} -
\frac{4}{3} \cdot 5^{-5/n} \right) \right\}~.
\end{eqnarray}

In the limit $n \rightarrow \infty$, Eq.\
(\ref{1stOrderLTBbackreactionResult}) gives
\begin{equation}\label{1stOrderLTBbackreactionLargen}
\mathcal{Q} = \frac{4}{135} t_0^{-2} \left(\frac{r_0}{R}
\right)^3\left(\frac{r_0}{t_0}\right)^2
 ~,
\end{equation}
showing that although both the average shear and the variance of
the expansion rate diverge as $n \rightarrow \infty$, the
backreaction, given by their difference, has a finite limit. Note
that, in the validity region of Fig.\ \ref{approxi1}, the
backreaction (\ref{1stOrderLTBbackreactionResult}) is
qualitatively similar for all values of $n$, but we consider the
step function limit to make comparison with the model that
consists of two disjoint FRW solutions in Sect.\ \ref{FRWmodels}.

\subsection{Comparison with a model that consists of two disjoint
FRW solutions}\label{FRWmodels}

Let us compare the exact solution of the void-wall configuration
to a model where the walls and voids are treated as separate FRW
regions, following the idea outlined in Ref.\
\cite{Rasanen:2006kp}. This comparison directly addresses the role
of shear since the shear is by construction zero in the disjoint
FRW approximation.

For the void or the $\Omega=0$ region we use the Milne solution
for which
\begin{equation}\label{Milne}
\theta_v = \frac{3}{t}~,
\end{equation}
and for the wall or the $\Omega=1$ region we use the Einstein de
Sitter or EdS solution for which
\begin{equation}\label{EDS}
\theta_w = \frac{2}{t}~.
\end{equation}
Since the shear vanishes in the FRW models, the backreaction is
given just by the variance
\begin{equation}\label{QFRW}
\mathcal{Q}_{{\rm FRW}} = \frac{2}{3} \left( \langle \theta^2
\rangle - \langle \theta \rangle^2 \right) ~,
\end{equation}
where
\begin{equation}\label{QFRW2}
\langle \theta^2 \rangle = \frac{V_w \theta_w^2 + V_v
\theta_v^2}{V_w + V_v}
\end{equation}
and
\begin{equation}\label{QFRW3}
\langle \theta \rangle = \frac{V_w \theta_w + V_v \theta_v}{V_w +
V_v}~.
\end{equation}
By inserting Eqs.\ (\ref{Milne}) and (\ref{EDS}) in Eqs.\
(\ref{QFRW2}) and (\ref{QFRW3}), we obtain
\begin{equation}\label{QFRW4}
\mathcal{Q}_{{\rm FRW}} = \frac{2}{3} t_0^{-2} f_v (1- f_v)~,
\end{equation}
where $f_v$ is the void fraction:
\begin{equation}\label{voidfraction}
f_v \equiv \frac{V_v}{V_v+V_w}~.
\end{equation}
For realistic-sized voids, $r_0 \ll t_0$, we have
\begin{equation}\label{}
V_v = \frac{4}{3} \pi r_0^3 \left[ 1 + \mathcal{O}
\left(\frac{r_0}{t_0}\right)^2 \right] \simeq \frac{4}{3} \pi
r_0^3~,
\end{equation}
while the wall is spatially flat, so Eq.\ (\ref{QFRW4}) can also
be written as
\begin{equation}\label{QFRW5}
\mathcal{Q}_{{\rm FRW}} = \frac{2}{3} t_0^{-2} \left(
\frac{r_0}{R} \right)^3 \left[ 1- \left( \frac{r_0}{R} \right)^3
\right]~,
\end{equation}
For the void fraction value $f_v=1/2$ or $R= 2^{1/3} r_0$, the
backreaction (\ref{QFRW5}) gets its maximum value:
\begin{equation}\label{QFRW6}
\mathcal{Q}_{{\rm FRW}} = \frac{1}{6} t_0^{-2} ~.
\end{equation}
We can compare this to the LTB result by setting $R= 2^{1/3} r_0$
likewise in Eq.\ (\ref{1stOrderLTBbackreactionLargen}):
\begin{equation}\label{QFRW7}
\mathcal{Q} = \frac{2}{135} t_0^{-2} \left( \frac{r_0}{t_0}
\right)^2 ~,
\end{equation}
implying the discrepancy:
\begin{equation}\label{QFRW8}
\mathcal{Q} = \frac{4}{45} \left( \frac{r_0}{t_0} \right)^2
\mathcal{Q}_{{\rm FRW}}~.
\end{equation}
The result (\ref{QFRW8}) demonstrates how important it can be to
take into account the shear: for a realistic-sized void with $r_0
=0.01 t_0$, the FRW approximation overestimates the backreaction
by the tremendous factor of $10^{5}$. The suppressive factor
$(r_0/t_0)^2$ appears to be consistent with the results from
perturbative analysis in Refs.\
\cite{Gruzinov:2006nk,Baumann:2010tm}.

\subsection{Uncompensated voids}\label{uncompensated}

\begin{figure}[tbh]
\begin{center}
\includegraphics[width=10.0cm]{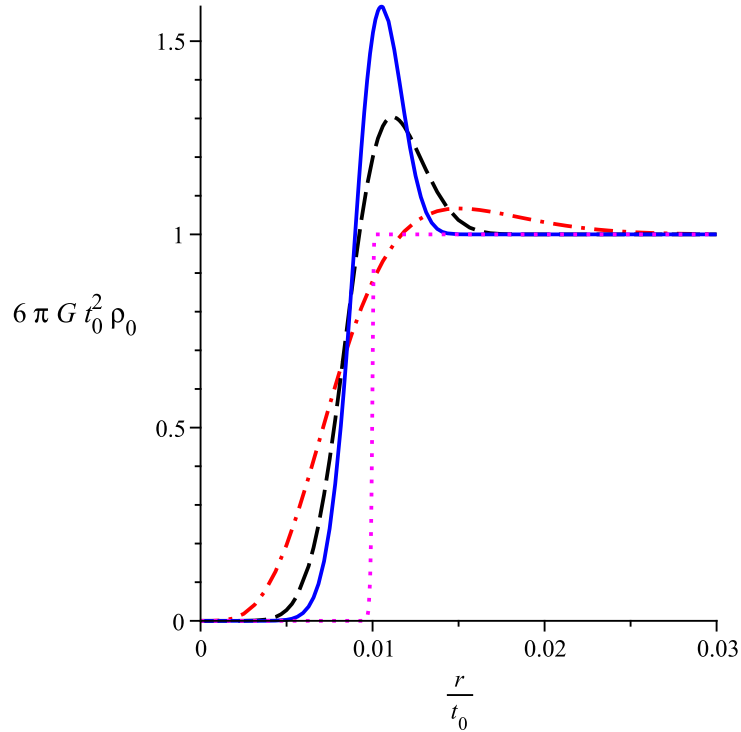}
\caption{Matter density as a function of $r$ for the profile
(\protect \ref{rhoexp}) with $n \rightarrow \infty$ (magenta
dotted curve) and the profile (\protect \ref{OmegaLTB}) with three
different values of $n$: $n=2$ (red dash dotted curve), $n=4$
(black dashed curve), $n=6$ (blue solid curve). All profiles in
this figure have $r_0 = 0.01 t_0$.} \label{rho1}
\end{center}
\end{figure}

The void-wall profiles (\ref{OmegaLTB}) considered in Sects.\
\ref{sier}--\ref{FRWmodels} have a compensating overdense peak in
the matter density between the void and the wall regions; see
Fig.\ \ref{rho1}. The peak is required to make the wall (very
close to) a flat FRW region, as can be seen from Eqs.\
(\ref{rooOomega}) and (\ref{LTBomega3}). In this section, we
consider profiles without the peak to see whether it plays an
important role in suppressing the backreaction relative to the FRW
value as in Eq.\ (\ref{QFRW8}).

Instead of specifying $\Omega_0(r)$ and $t_0$, a simultaneous Big
Bang LTB model can alternatively be parameterized by the physical
matter density profile $\rho_0(r)$ on a spatial hypersurface at
$t=t_0$. In this way, we can avoid the compensating overdensity by
choosing a monotonically increasing density profile such as
\begin{equation}\label{rhoexp}
\rho_0 (r) = \rho_{{\rm{crit}}} (\infty) \left( 1 - e^{-(r/r_0)^n}
\right) ~,
\end{equation}
where $\rho_{{\rm{crit}}} (\infty) = 1/6 \pi G t_0^2$.
Substituting Eqs.\ (\ref{H0-OmegaRelationAppr}) and (\ref{rhoexp})
in  Eq.\ (\ref{LTBomega}), we obtain
\begin{equation}\label{omegarhoexpn}
\Omega_0(r) = \frac{9}{4} \left( 1 - \sqrt{ 1 -
\frac{8}{3\sqrt{3}} \sqrt{ \sum_{l=1}^{\infty} \frac{(-1)^{l+1}
\left(r/r_0\right)^{nl}}{l! (nl+3)} } } \right)^2~,
\end{equation}
which, by virtue of Eq.\ (\ref{H0-OmegaRelationAppr}), implies
\begin{equation}\label{H0rhoexpn}
H_0(r) = \frac{t_0^{-1}}{2} \left( 1 + \sqrt{1-
\frac{8}{3\sqrt{3}} \sqrt{ \sum_{l=1}^{\infty} \frac{(-1)^{l+1}
\left(r/r_0\right)^{nl}}{l! (nl+3)} }} \right) ~.
\end{equation}
In the step function limit $n \rightarrow \infty$, Eqs.\
(\ref{omegarhoexpn}) and (\ref{H0rhoexpn}) become
\begin{equation}\label{omegarhostep}
\Omega_0(r) = \frac{9}{4} \left( 1 - \sqrt{ 1 - \frac{8}{9}
\sqrt{1-\left( \frac{r_0}{r} \right)^3} } \right)^2 \Theta(r-r_0)
\end{equation}
and
\begin{equation}\label{H0rhostep}
H_0(r) = t_0^{-1} \left[1 -  \frac{1}{2} \left( 1 - \sqrt{1 -
\frac{8}{9} \sqrt{1-\left( \frac{r_0}{r} \right)^3}  } \right)
\Theta (r-r_0)
 \right]~,
\end{equation}
where $\Theta(r-r_0) = \lim_{n \rightarrow \infty} (1 -
e^{-(r/r_0)^n})$ stands for the Heaviside step function.

The backreaction (\ref{defBackreaction}) cannot be integrated
analytically for the profile (\ref{rhoexp}). Instead, performing
the integrals numerically for the parameter values $(r_0/R)^3=1/2$
and $r_0=0.01t_0$, we obtain
\begin{equation}\label{RhostepQ}
\mathcal{Q} = 1.6 \cdot 10^{-6} t_0^{-2}~.
\end{equation}
This differs very little from the case with the step function in
$\Omega_0(r)$ and the compensating overdensity in the matter
distribution: for comparison, substituting the values
$(r_0/R)^3=1/2$ and $r_0=0.01t_0$ in Eq.\ (\ref{QFRW7}) yields
\begin{equation}\label{OmegastepQ}
\mathcal{Q} = 1.5 \cdot 10^{-6} t_0^{-2}~.
\end{equation}
The near equality of the results (\ref{RhostepQ}) and
(\ref{OmegastepQ}) implies that the suppression factor appearing
in Eq.\ (\ref{QFRW8}) is not due to the overdense peak between the
wall and the void regions. Furthermore, keeping the void-wall
volume ratio fixed to $(r_0/R)^3=1/2$ and varying $r_0$ shows that
the dependence of the backreaction on $r_0$ is very similar to
Eq.\ (\ref{QFRW7}), i.e. $\mathcal{Q} \propto r_0^2$ also for the
profile (\ref{rhoexp}).

Extrapolating the result from a void-wall pair to the network of
voids is harder for the uncompensated voids than for the
compensated ones. The reason is that, without the compensating
overdensity, the solution approaches FRW metric slower: although
the matter density is constant outside the void ($r
> r_0$) for the profile (\ref{rhoexp}) in the limit $n \rightarrow
\infty$, the spatial curvature becomes (nearly) constant only much
further ($r \gtrsim 10 r_0$) from the void. It thus appears that
in order to match together uncompensated voids in the simple
fashion explained in Sect.\ \ref{model}, the separation between
the voids must be so large that only a few voids fit inside the
horizon. Therefore, more sophisticated junction conditions need to
be applied in joining together uncompensated voids to obtain a
global void-fraction more consistent with observations such as the
ones quoted in Ref.\ \cite{Hoyle:2003hc}.

\section{Conclusions}\label{konkluusiot}

We have considered the role of shear in the cosmological
backreaction problem using the spherically symmetric LTB dust
solution. The LTB models utilized in Sects.\
\ref{sier}--\ref{FRWmodels} have a close to simultaneous Big Bang
with the inhomogeneity profile (\ref{OmegaLTB}) that at $t=t_0$
interpolates between a void region (where $\Omega_0 \simeq 0$ and
$H_0 t_0 \simeq 1$) and a wall region (where $\Omega_0 \simeq 1$
and $H_0 t_0 \simeq 2/3$) with the parameters $r_0$ and $n$
describing the size of the void and the sharpness of the
transition between the two regions, respectively.

For realistic-sized voids ($r_0 \ll t_0$) with a fixed void-wall
volume ratio, the results of Sect.\ \ref{sier} show that the
volume-average shear is independent of $r_0$ or the size of the
void. This, along with the fact that the different void-wall
solutions naturally join together in the EdS-like wall region,
implies that the volume-average shear found for a single void-wall
pair actually applies also for a \emph{network} of voids of
different size. Moreover, we found that for values $n\gtrsim 10$
both the volume-average shear and the variance of the expansion
rate grow proportional to $n$, hence diverging in the limit $n
\rightarrow \infty$. However, as shown in Sect.\
\ref{Sbackreaction}, the backreaction, given by the difference of
the variance and the shear, has a finite limiting value for $n
\rightarrow \infty$.

In Sect.\ \ref{FRWmodels}, to estimate the role of the shear in
the backreaction, we compared the LTB result for the backreaction
in the limit $n \rightarrow \infty$ to a simplified model that
ignores the shear by approximating the void-wall configuration
with two disjoint FRW solutions: $\Omega_0=0$ or the Milne
solution for the void and $\Omega_0=1$ or the EdS solution for the
wall. The comparison shows that the backreaction obtained using
the exact LTB solution is suppressed by at least a factor of
$(r_0/t_0)^2$ relative to the value obtained from the model made
up of the separate FRW solutions, hence implying that the shear
plays a crucial role here. In Sect.\ \ref{uncompensated}, we
demonstrated that the suppression is not due to a compensating
overdensity between the void and the wall regions.

The LTB-based void-wall models considered here contain various
simplifications that may misestimate the backreaction of the real
universe. Firstly, the models lack globally significant spatial
curvature which is known to be important for large backreaction
\cite{Buchert:2007ik}. Large negative spatial curvature can arise
in the late universe if its volume becomes dominated by
uncompensated voids. This can be seen by realizing that a dense
enough network of uncompensated voids is dynamically similar to a
huge negatively curved void of size $r_0/t_0 \simeq
\mathcal{O}(1)$. Another issue is the spherical symmetry of the
voids which is obviously broken in the real universe. Finally, the
models considered here do not contain collapsing regions, which
have been suggested to play an essential role in the average
dynamics of the universe \cite{Rasanen:2006kp,Rasanen:2008it}.

Besides the dynamical backreaction (\ref{defBackreaction}) studied
in this work, the cosmological structures may have other effects
on the observations that are not captured by a spatial averaging
procedure. Proposed examples include effects on the propagation of
light
\cite{Mattsson:2007tj,Rasanen:2008be,Enqvist:2009hn,Clifton:2009jw,Kainulainen:2009dw,Rasanen:2009uw}
and effects due to our non-average location in the universe
\cite{Wiltshire:2007zj,Wiltshire:2007jk,Wiltshire:2008sg,Wiltshire:2009db}.
Therefore, small dynamical backreaction does not alone imply that
the effects of the inhomogeneities on the observations were small
or insignificant.

\acknowledgments{We thank David Wiltshire and Charles Hellaby for
helpful discussions and Tomi Koivisto for useful comments. This
work was supported by the Marsden fund of the Royal Society of New
Zealand. MM is supported by the Graduate School in Particle and
Nuclear Physics (GRASPANP) and acknowledges The Magnus Ehrnrooth
foundation for supporting her visit in the University of
Canterbury, where the major part of this work was done, and David
Wiltshire for hospitality. TM acknowledges The Emil Aaltonen
foundation for support.}

\end{document}